\documentclass[english,preprint,letter,showpacs,preprintnumbers,amssymb,secnumarabic]{revtex4-1}
\usepackage[T1]{fontenc}
\usepackage[latin9]{inputenc}
\usepackage{units}
\usepackage{amstext}
\usepackage{graphicx}

\makeatletter

\newcommand{\binom}[2]{{#1 \choose #2}}

\newcommand{\lyxdot}{.}

 \@ifundefined{textcolor}{}
 {%
   \definecolor{BLACK}{gray}{0}
   \definecolor{WHITE}{gray}{1}
   \definecolor{RED}{rgb}{1,0,0}
   \definecolor{GREEN}{rgb}{0,1,0}
   \definecolor{BLUE}{rgb}{0,0,1}
   \definecolor{CYAN}{cmyk}{1,0,0,0}
   \definecolor{MAGENTA}{cmyk}{0,1,0,0}
   \definecolor{YELLOW}{cmyk}{0,0,1,0}
 }

\makeatother

\usepackage{babel}
\begin{document}

\title{Strongly Intensive Cumulants: Fluctuation Measures for Systems With
Incompletely Constrained Volumes}

\author{Evan Sangaline}

\affiliation{National Superconducting Cyclotron Laboratory, Michigan State University,
East Lansing, Michigan 48824, USA}
\begin{abstract}
The cumulants of thermal variables are of general interest in physics
due to their extensivity and their correspondence with susceptibilities.
They become especially significant near critical points of phase transitions
where they diverge along with the correlation length. Cumulant measurements
have been used extensively within the field of heavy-ion physics,
principally as tools in the search for a hypothetical QCD critical
point along the transition between hadronic matter and QGP. The volume
of individual heavy-ion collisions can be only partially constrained
and, as a result, cumulant measurements are significantly biased by
the limited volume resolution. We propose a class of moments called
strongly intensive cumulants which can be accurately measured in the
presence of unconstrained volume fluctuations. Additionally, they
share the same direct relationship with susceptibilities as cumulants
in many cases.
\end{abstract}
\maketitle

\section{Introduction}

The development of classes of moments that are either invariant or
have clear scaling properties under various operations has been an
active area of research for well over a century. The initial work
on cumulants themselves predated the development of statistical partition
functions and focused on their scaling properties under affine transformations
\citep{thiele}. They were originally called half-invariants due to
these scaling properties and were later renamed cumulants due to their
additivity under convolution \citep{fisher}. In the field of computer
vision, two-dimensional moments which are invariant under operations
such as rotation, scaling, and translation have played a central role
in pattern recognition \citep{hu-invariants,image-invariants}. Within
the field of particle physics, the $R$ fluctuation measures were
constructed as ratios of factorial moments such that detector efficiencies
would cancel resulting in moments invariant under binomial efficiency
losses \citep{R-76,R-75}. 

More recently, the $\Delta$ and $\Sigma$ observables were proposed
to address the issue of measurement biases resulting from the poor
constraint of volume in heavy-ion collisions \citep{merick-1992,merick-2013,merik=2011}.
These quantities are two-dimensional second-order moments which are
constructed in such a way that the volume fluctuation terms cancel.
The authors coined the term \textit{strongly intensive} to describe
these observables because they are not only independent of the volume
of the system but also of the distribution of volume within an ensemble.
They have been used effectively in fluctuation analyses \citep{delta-sigma-1,delta-sigma-2},
but their physical meanings are obscure relative to those of cumulants
and they only measure second-order fluctuations.

In this work, we present a new set of statistical quantities which
we call \textit{strongly intensive cumulants}. These quantities are
invariant under both convolution and mixing with distributions sharing
the same strongly intensive cumulants. A direct result of this is
that they can accurately be measured experimentally in situations
where the volume is not well constrained. These quantities are directly
related to cumulants when certain conditions are met and therefore,
in physical systems, to thermodynamic susceptibilities.

The strongly intensive cumulants are of particular interest in heavy-ion
collisions where cumulant measurements of conserved charges have been
proposed as a signature for critical point fluctuations \citep{Stephanov2009}.
The experimental determination of these cumulants has been a focus
of the RHIC beam energy scan \citep{net_charge_star,net_proton_star}
and the published measurements have been discussed in the context
of critical fluctuations as well as compared to lattice calculations
in order to determine the temperature and chemical potentials at chemical
freeze-out \citep{swagato}. It has been shown that these measurements
depend greatly on the method used to constrain collision volumes which
calls into question the validity of any physics conclusions drawn
from the results \citep{Westfall2014}.

To demonstrate the utility and efficacy of the strongly intensive
cumulants we present an analysis of net proton fluctuations in simulated
heavy-ion collisions. This analysis makes clear the issues with current
analysis techniques and illustrates how the strongly intensive cumulants
resolve them. In light of this, we propose that the strongly intensive
cumulants listed in Eq. (\ref{eq:first-four-strongly intensive cumulants})
be used as drop-in replacements in future heavy-ion cumulant analyses.

\section{motivation}

We begin by briefly reviewing the generating function formalism with
respect to statistical moments, cumulants, and their relationships
to volume fluctuations. Let $X=\left(X_{1},X_{2},\dots,X_{n}\right)$
be the components of a random vector. The moment-generating function
is then defined to be

\begin{eqnarray*}
\phi\left(\xi\right) & \equiv & \left\langle e^{\xi_{i}X_{i}}\right\rangle _{X}\\
 & = & \sum_{r_{1=0}}^{\infty}\sum_{r_{2=0}}^{\infty}\cdots\sum_{r_{n=0}}^{\infty}\frac{\mu_{r_{1},r_{2},\dots,r_{n}}}{r_{1}!r_{2}!\cdots r_{n}!}\xi_{1}^{r_{1}}\xi_{2}^{r_{2}}\cdots\xi_{n}^{r_{n}}
\end{eqnarray*}
where the $\left\langle \cdots\right\rangle _{X}$ notation denotes
the expectation value over the distribution of $X$ and the repeated
index $i$ is implicitly summed over according to Einstein summation
notation. The coefficients of the Taylor series correspond to the
moments of the distribution and can be recovered from the generating
function by taking derivatives and setting $\xi=0$. Defining the
operator $D_{i}$ to be $\nicefrac{\partial}{\partial\xi_{i}}$, we
can see that 
\begin{eqnarray*}
\mu_{r_{1},r_{2},\dots,r_{n}} & = & \left[D_{1}^{r_{1}}D_{2}^{r_{2}}\cdots D_{n}^{r_{n}}\left(\phi\right)\right]_{\xi=0}\\
 & = & \left\langle X_{1}^{r_{1}}X_{2}^{r_{2}}\cdots X_{n}^{r_{n}}\right\rangle _{X}
\end{eqnarray*}
as we would expect. This approach of recovering coefficients from
the generating function can be applied in the same way for both the
standard cumulants and the strongly intensive cumulants that we will
introduce later.

The generating function for the cumulants is then defined in terms
of the moment-generating function as
\begin{eqnarray*}
\psi\left(\xi\right) & \equiv & \ln\phi\left(\xi\right)\\
 & = & \sum_{r_{1=0}}^{\infty}\sum_{r_{2=0}}^{\infty}\cdots\sum_{r_{n=0}}^{\infty}\frac{\kappa_{r_{1},r_{2},\dots,r_{n}}}{r_{1}!r_{2}!\cdots r_{n}!}\xi_{1}^{r_{1}}\xi_{2}^{r_{2}}\cdots\xi_{n}^{r_{n}}
\end{eqnarray*}
where the coefficients $\kappa_{i,j,\dots,k}$ are the cumulants and
can be expressed in terms of the moments by matching terms in the
Taylor series, either combinatorially or using recursion relations
\citep{recursive_cumulants}.

The utility of the cumulants becomes clear when we consider the convolution
of probability distributions. If we take two independent random vectors,
$X$ and $Y$, then the distribution of $Z=X+Y$ is given by the convolution
of their respective distributions. The moment-generating function
of $Z$ is then given by $\phi_{Z}\left(\xi\right)=\left\langle e^{\xi_{i}\left(X+Y\right)_{i}}\right\rangle _{X,Y}$
which, due to the independence of $X$ and $Y$, can be factored as
$\phi_{Z}\left(\xi\right)=\left\langle e^{\xi_{i}X_{i}}\right\rangle _{X}\left\langle e^{\xi_{i}Y_{i}}\right\rangle _{Y}$.
The cumulant-generating function is then simply $\psi_{Z}\left(\xi\right)=\ln\phi_{Z}\left(\xi\right)=\ln\phi_{X}\left(\xi\right)+\ln\phi_{Y}\left(\xi\right)$
which is precisely the sum of the cumulant-generating functions for
$X$ and $Y$. Thus, we see that the cumulants of two probability
distributions are additive under convolution, a fact that led to their
current name.

This additivity property of cumulants is closely related to their
utility in physics. If we consider two volumes of matter that are
each in the same thermodynamic state then the distributions of any
total quantities (e.g. net charges, total energy) for the combined
volume will be given by the convolution of their distributions for
each of the two independent volumes. The cumulants of these distributions
will necessarily be extensive and, after scaling by the volume, will
give intrinsic quantities determined by the thermodynamic state of
the matter. By relating the partition function to a moment-generating
function we can see that the coefficients of the corresponding cumulant-generating
function then encode how the mean values of total quantities change
with respect to state variables like energy density or chemical potential
\citep{relating-cumulants-physics}.

The determination of these coefficients is of general interest in
physics but there is an experimental limitation that makes it difficult.
If the volume of an ensemble of systems cannot be perfectly constrained
then the cumulants of the distributions depend strongly on the distribution
over volume within the ensemble. To see this we define an intrinsic
generating function such that

\begin{eqnarray}
\psi^{\prime}\left(\xi\right) & \equiv & \psi\left(\xi\right)/V\label{eq:volume-factored}\\
 & = & \sum_{r_{1=0}}^{\infty}\sum_{r_{2=0}}^{\infty}\cdots\sum_{r_{n=0}}^{\infty}\frac{\kappa_{r_{1},r_{2},\dots,r_{n}}^{\prime}}{r_{1}!r_{2}!\cdots r_{n}!}\xi_{1}^{r_{1}}\xi_{2}^{r_{2}}\cdots\xi_{n}^{r_{n}}\nonumber 
\end{eqnarray}
where the primes indicate independence from the volume. We use the
term volume here loosely; it really corresponds to any measure with
which the cumulants scale linearly. In actuality, $V$ can depend
on the temperature, the energy density, and other quantities in addition
to the volume. The important point in the context of this discussion
is simply that this quantity can be factored out from the cumulant-generating
function, as done in Eq. (\ref{eq:volume-factored}). 

The moment-generating function can then be expressed as
\begin{eqnarray*}
\phi\left(\xi\right) & = & e^{V\psi^{\prime}\left(\xi\right)}\\
 & = & 1+V\psi^{\prime}\left(\xi\right)+\frac{1}{2!}V^{2}\psi^{\prime}\left(\xi\right)^{2}+\cdots
\end{eqnarray*}
with an explicit volume dependence. In any experimental context, the
moments that are measured are those of a mixed distribution over an
ensemble of volumes. The measured cumulants will then be described
by the generating function
\begin{eqnarray}
\psi\left(\xi\right) & = & \ln\left\langle e^{V\psi^{\prime}\left(\xi\right)}\right\rangle _{V}\label{eq:cumulant-gf-with-volume}\\
 & = & \ln\left(1+\left\langle V\right\rangle _{V}\psi^{\prime}\left(\xi\right)+\frac{1}{2!}\left\langle V^{2}\right\rangle _{V}\psi^{\prime}\left(\xi\right)^{2}+\cdots\right)\nonumber 
\end{eqnarray}
 which can easily be seen to not equal $\left\langle V\right\rangle _{V}\psi^{\prime}\left(\xi\right)$
unless $V$ is fixed at a single value.

The measured cumulants instead depend on the distribution of the volume
in a straight-forward way \citep{volume_fluctuations}. This relationship
can be made more clear by considering the simple example of the variance
of a single variable. By expanding the logarithm in Eq. (\ref{eq:cumulant-gf-with-volume})
and matching terms we find that the measured variance would be
\begin{equation}
\kappa_{2}=\left\langle V\right\rangle _{V}\kappa_{2}^{\prime}+\left(\left\langle V^{2}\right\rangle _{V}-\left\langle V\right\rangle _{V}^{2}\right)\left(\kappa_{1}^{\prime}\right)^{2}\label{eq:second-order-fluctuations}
\end{equation}
in the presence of volume fluctuations. If the volume is constrained
to a single value then the variance of $V$ goes to zero and the relationship
reduces to $\kappa_{2}=\left\langle V\right\rangle _{V}\kappa_{2}^{\prime}$,
the quantity that one would truly want to measure. The measured variance
will be artificially high for any other distribution of volume due
to the contribution of the second term. The most extreme example of
this is a situation where $\kappa_{2}^{\prime}=0$, which would be
approximately true for the total number of atoms in a crystal lattice.
In this scenario, the first term on the right hand side of Eq. (\ref{eq:second-order-fluctuations})
vanishes and the measured variance would be directly proportional
to the variance of the volume in the measurement ensemble.

It may seem as though the fluctuation terms could be subtracted off
from measurements such that the cumulants$\left\langle V\right\rangle _{V}\kappa_{i}^{\prime}$
could be directly determined. This is true in theory, but the measurements
cannot be corrected for without an exact knowledge of the volume distribution.
The precise shape of this distribution is typically not known in practice
and so this approach is not applicable. Instead, measurements tend
to be made without any attempt at corrections and, as a result, they
are biased in poorly understood ways due to the contributions from
volume fluctuations.

\section{Derivation}

We introduce here a set of statistical quantities called the strongly
intensive cumulants. The goal is to construct a set of non-trivial
statistical quantities that can be measured in physical systems without
any dependence on the volume distribution. Formally, this is equivalent
to saying that the quantities are invariant under both convolution
and mixing of distributions sharing the same strongly intensive cumulants.
We will begin by defining the strongly intensive cumulants and then
prove that they satisfy the desired properties.

Their generating function, $\psi^{*}$, is defined in terms of the
partial differential equation
\begin{equation}
D_{u}\left(\psi^{*}\right)=\frac{D_{u}\left(\phi\right)}{D_{v}\left(\phi\right)}\label{eq:strongly intensive cumulant-definition}
\end{equation}
where the choice of $u$ and $v$ determines the exact flavor of the
generating function. We assume from here foreword that the $u$ and
$v$ indices are $1$ and $n$, respectively. The components of $X$
can always be rearranged such that this is the case, with the trivial
exception of $u=v$. Without loss of generality, we can assume this
has been done. 

The choice of the first and last components of $X$ is arbitrary,
but we will see in Sec. \ref{sec:Relation-to-Other} that it makes
no difference which component comes first with an appropriate choice
of $X_{n}$. In this situation, $X_{n}$ serves as a measure of the
volume and is independent of the quantities that one is primarily
interested in. The general case is more subtle but thinking of a physical
situation where $X_{n}$ is a noisy volume measurement and the choice
of $X_{1}$ is arbitrary can be helpful in understanding how we proceed.

The strongly intensive cumulants, $\kappa_{r_{1},r_{2},\dots,r_{2}}$,
then correspond to the Taylor series coefficients of

\begin{eqnarray*}
\psi^{*}\left(\xi\right) & = & \sum_{r_{1}=0}^{\infty}\sum_{r_{2}=0}^{\infty}\cdots\sum_{r_{n}=0}^{\infty}\frac{\kappa_{r_{1},r_{2},\dots,r_{n}}^{*}}{r_{1}!r_{2}!\cdots r_{n}!}\xi_{1}^{r_{1}}\xi_{2}^{r_{2}}\cdots\xi_{n}^{r_{n}}
\end{eqnarray*}
in the same way that cumulants are defined by the coefficients of
$\psi$. The $\kappa_{0,r_{2},r_{3},\dots,r_{n}}^{*}$ terms in this
series are, as of yet, undetermined because they correspond to the
integration constants obtained when integrating Eq. (\ref{eq:strongly intensive cumulant-definition})
with respect to $\xi_{1}$. We choose to define these terms to be
$\kappa_{0,r_{2},r_{3},\dots,r_{n}}^{*}=\kappa_{r_{2},r_{3},\dots,r_{n}}^{*}$,
the strongly intensive cumulant obtained when the first component
of $\xi$ is dropped such that $\xi\rightarrow\left(\xi_{2},\xi_{3},\dots,\xi_{n}\right)$.
This process can be applied repeatedly until the first index is nonzero
or, if all indices are zero, $\kappa_{0,0,\dots,0}^{*}=0$. This definition
allows us to prove properties for $D_{1}\left(\psi^{*}\right)$ and
have them apply more generally to $\psi^{*}$ because the missing
$\kappa_{0,r_{2},r_{3},\dots,r_{n}}^{*}$ terms correspond to coefficients
of $D_{1}\left(\psi^{*}\right)$ for a different choice of $\xi$.

First, we'll show explicitly that the strongly intensive cumulants
are completely independent of the distribution over volume in a physical
system. This can be proved very simply by substituting the volume
mixed moment-generating function from earlier into Eq. (\ref{eq:strongly intensive cumulant-definition})
\begin{eqnarray}
D_{1}\left(\psi^{*}\right) & = & \frac{D_{1}\left(\left\langle e^{V\psi^{\prime}\left(\xi\right)}\right\rangle _{V}\right)}{D_{n}\left(\left\langle e^{V\psi^{\prime}\left(\xi\right)}\right\rangle _{V}\right)}\label{eq:proof-of-volume-canceling}\\
 & = & \frac{\left\langle Ve^{V\psi^{\prime}\left(\xi\right)}\right\rangle _{V}D_{1}\left(\psi^{\prime}\right)}{\left\langle Ve^{V\psi^{\prime}\left(\xi\right)}\right\rangle _{V}D_{n}\left(\psi^{\prime}\right)}=\frac{D_{1}\left(\psi^{\prime}\right)}{D_{n}\left(\psi^{\prime}\right)}\nonumber 
\end{eqnarray}
and canceling the volume dependent terms. This proof of their strongly
intensive property is straightforward, but it is not the most clear
way to demonstrate why this happens. By exploring the properties of
the strongly intensive cumulants under convolution and mixing it will
become more clear.

We will first consider convolutions, whereby a new random vector $Z$
is constructed as the sum of two other random vectors $X$ and $Y$.
For physical systems, this operation can be viewed as constructing
a larger volume out of two smaller volumes. Given that $Z=X+Y$, we
find that $\psi_{Z}\left(\xi\right)=\psi_{X}\left(\xi\right)+\psi_{Y}\left(\xi\right)$
or, equivalently, that $\phi_{Z}\left(\xi\right)=\phi_{X}\left(\xi\right)\phi_{Y}\left(\xi\right)$.
Plugging this into Eq. (\ref{eq:strongly intensive cumulant-definition})
we find that

\begin{equation}
D_{1}\left(\psi_{Z}^{*}\right)=\frac{D_{n}\left(\psi_{X}\right)D_{1}\left(\psi_{X}^{*}\right)+D_{n}\left(\psi_{Y}\right)D_{1}\left(\psi_{Y}^{*}\right)}{D_{n}\left(\psi_{X}\right)+D_{n}\left(\psi_{Y}\right)}\label{eq:convolution equation}
\end{equation}
which can be seen as a $\xi$-dependent weighted average of the differential
equations defining the strongly intensive cumulant generating functions
of $X$ and $Y$.

Now let us consider the case that $X$ and $Y$ have the same set
of strongly intensive cumulants. This implies that $D_{1}\left(\psi_{X}^{*}\right)=D_{1}\left(\psi_{Y}^{*}\right)$
which simplifies Eq. (\ref{eq:convolution equation}) to
\begin{eqnarray}
D_{1}\left(\psi_{Z}^{*}\right) & = & \frac{D_{n}\left(\psi_{X}\right)D_{1}\left(\psi_{X}^{*}\right)+D_{n}\left(\psi_{Y}\right)D_{1}\left(\psi_{X}^{*}\right)}{D_{n}\left(\psi_{X}\right)+D_{n}\left(\psi_{Y}\right)}\nonumber \\
 & = & \frac{D_{n}\left(\psi_{X}\right)+D_{n}\left(\psi_{Y}\right)}{D_{n}\left(\psi_{X}\right)+D_{n}\left(\psi_{Y}\right)}D_{1}\left(\psi_{X}^{*}\right)\label{eq:invariant-under-convolution}\\
 & = & D_{1}\left(\psi_{X}^{*}\right)=D_{1}\left(\psi_{Y}^{*}\right)\nonumber 
\end{eqnarray}
showing that the strongly intensive cumulants are invariant under
the convolution of distributions with identical strongly intensive
cumulants. This demonstrates the intensive property of the strongly
intensive cumulants and how it emerges from the way they combine under
convolution.

The situation for mixing distributions very closely parallels that
for convolving them. If convolution can be thought of as an operation
for constructing new volumes then distribution mixing can be thought
of as the operation of combining different volumes into an ensemble.
If we imagine that half of the time we choose a random vector $Z$
according to $P_{X}\left(Z\right)$ and half of the time we choose
it from $P_{Y}\left(Z\right)$ then the probability for the mixed
distribution $Z$ is given by $P_{Z}\left(Z\right)=\frac{1}{2}\left(P_{X}\left(Z\right)+P_{Y}(Z)\right)$.
We could have chosen a ratio other than $\frac{1}{2}$ but the extension
to a mixing parameter is trivial. The resulting moment-generating
function is, like the probability distribution, the arithmetic average
of those for the separate distributions: $\phi_{Z}\left(\xi\right)=\frac{1}{2}\left(\phi_{X}\left(\xi\right)+\phi_{Y}\left(\xi\right)\right)$.
This results in a strongly intensive cumulant generating function
of 
\begin{equation}
D_{1}\left(\psi_{Z}^{*}\right)=\frac{D_{n}\left(\phi_{X}\right)D_{1}\left(\psi_{X}^{*}\right)+D_{n}\left(\phi_{Y}\right)D_{1}\left(\psi_{Y}^{*}\right)}{D_{n}\left(\phi_{X}\right)+D_{n}\left(\phi_{Y}\right)}\label{eq:mixing-equation}
\end{equation}
which is again a weighted average of the two independent differential
equations.

It is important to note that Eq. (\ref{eq:mixing-equation}) is identical
to Eq. (\ref{eq:convolution equation}) with $\psi$ replaced by $\phi$.
Although the strongly intensive cumulants do not combine as simply
as the moments under mixing or the cumulants under convolution, they
have the same relationship with the moments under mixing as they do
with the cumulants under convolution. Similarly, we find that
\[
D_{1}\left(\psi^{*}\right)=\frac{D_{1}\left(\phi\right)}{D_{n}\left(\phi\right)}=\frac{\nicefrac{1}{\phi}D_{1}\left(\phi\right)}{\nicefrac{1}{\phi}D_{n}\left(\phi\right)}=\frac{D_{1}\left(\ln\phi\right)}{D_{n}\left(\ln\phi\right)}=\frac{D_{1}\left(\psi\right)}{D_{n}\left(\psi\right)}
\]
showing that the strongly intensive cumulants are directly related
to the cumulants and moments of a distribution in the same way (i.e.
the expressions for the strongly intensive cumulants are unchanged
when all moments are replaced with cumulants or vice versa). The similarity
in the relationships that the strongly intensive cumulants have with
both the moments and the cumulants is at the core of why they exhibit
strongly intensive behavior.

It follows from Eq.s (\ref{eq:convolution equation}) and (\ref{eq:mixing-equation})
that the strongly intensive cumulants are invariant under mixing and
convolution of distributions with identical strongly intensive cumulants.
This was shown explicitly for convolution but the same proof employed
in Eq. (\ref{eq:invariant-under-convolution}) applies to mixing when
Eq. (\ref{eq:mixing-equation}) is used as a starting point. The invariance
under convolution implies that they will be intensive, or independent
of volume, in a thermodynamic system. The combination of an ensemble
with many different volumes is, in effect, distribution mixing and
the strongly intensive cumulants will therefore be invariant under
this operation as well. This is equivalent to saying that the strongly
intensive cumulants can be measured over an ensemble of different
volumes without any dependence on the volume distribution. This line
of reasoning helps to illustrate why the volume terms were shown to
cancel in Eq. (\ref{eq:proof-of-volume-canceling}).

\section{Obtaining Expressions For The Strongly Intensive Cumulants}

We move now to the task of determining polynomial expressions for
the strongly intensive cumulants. Let $f\left(\xi\right)=D_{n}\left(\phi\right)$
and $g\left(\xi\right)=1/f\left(\xi\right)$ where the coefficients
in the Taylor series for $g\left(\xi\right)$ are given by $a_{r_{1},r_{2},\dots,r_{n}}/\left(r_{1}!r_{2}!\cdots r_{n}!\right)$.
We then find that 
\begin{eqnarray*}
0 & = & \left[D_{1}^{r_{1}}D_{2}^{r_{2}}\cdots D_{n}^{r_{n}}\left(f\times g\right)\right]_{\xi=0}\\
 & = & \sum_{i_{1}=0}^{r_{1}}\sum_{i_{2}=0}^{r_{2}}\cdots\sum_{i_{n}=0}^{r_{n}}\binom{r_{1}}{i_{1}}\binom{r_{2}}{i_{2}}\cdots\binom{r_{n}}{i_{n}}\\
 &  & \times a_{i_{1},i_{2},\dots,i_{n}}\mu_{r_{1}-i_{1},r_{2}-i_{2},\dots,r_{n-1}-i_{n-1},r_{n}+1-i_{n}}
\end{eqnarray*}
when at least one of $r_{1},r_{2},\dots,r_{n}$ are nonzero. We can
rearrange this to give a recursion equation

\begin{eqnarray*}
a_{r_{1},r_{2},...,r_{n}} & = & \underbrace{\sum_{i_{1}=0}^{r_{1}}\sum_{i_{2}=0}^{r_{2}}\cdots\sum_{i_{n}=0}^{r_{n}}}_{i_{1}\ne r_{1}\vee i_{2}\ne r_{2}\vee\cdots\vee i_{n}\ne r_{n}}\binom{r_{1}}{i_{1}}\binom{r_{2}}{i_{2}}\cdots\binom{r_{n}}{i_{n}}\\
 &  & \times a_{i_{1},i_{2},\dots,i_{n}}\frac{\mu_{r_{1}-i_{1},r_{2}-i_{2},\dots,r_{n-1}-i_{n-1},r_{n}+1-i_{n}}}{-\mu_{0,0,\dots,0,1}}
\end{eqnarray*}
with the starting point $a_{0,0,\dots,0}=g\left(0\right)=\nicefrac{1}{\mu_{0,0,\dots,0,1}}$.

Now we can express the strongly intensive cumulants in terms of these
coefficients in a similar manner. We find that

\begin{eqnarray*}
\kappa_{r_{1},r_{2},\dots,r_{n}}^{*} & = & \left[D_{1}^{r_{1}-1}D_{2}^{r_{2}}\cdots D_{n}^{r_{n}}\left(D_{1}\left(\phi\left(\xi\right)\right)g\left(\xi\right)\right)\right]\\
 & = & \sum_{i_{1}=0}^{r_{1}-1}\sum_{i_{2}=0}^{r_{2}}\cdots\sum_{i_{n}=0}^{r_{n}}\binom{r_{1}-1}{i_{1}}\binom{r_{2}}{i_{2}}\cdots\binom{r_{n}}{i_{n}}\\
 &  & \times a_{i_{1},i_{2},\dots,i_{n}}\mu_{r_{1}-i_{1},r_{2}-i_{2},\dots,r_{n}-i_{n}}
\end{eqnarray*}
which provides a full prescription for finding the strongly intensive
cumulants for $r_{1}>0$. If $r_{1}=0$ then the first component can
be removed, as described earlier. For the special case of $n=2$ and
$r_{2}=0$ this can be simplified to
\[
\kappa_{r}^{*}\equiv\kappa_{r,0}^{*}=\frac{1}{\mu_{0,1}}\left(\mu_{r,0}-\sum_{i=1}^{r-1}\binom{r-1}{i-1}\mu_{r-i,1}\kappa_{i}^{*}\right)
\]
where the second zero index has been dropped. This can be done without
ambiguity because the strongly intensive cumulants are trivial in
one dimension. This recursive equation can be used to find the explicit
expressions
\begin{eqnarray}
\kappa_{1}^{*} & = & \frac{\mu_{1,0}}{\mu_{0,1}}\nonumber \\
\kappa_{2}^{*} & = & \frac{\mu_{2,0}}{\mu_{0,1}}-\frac{\mu_{1,0}\mu_{1,1}}{\mu_{0,1}{}^{2}}\nonumber \\
\kappa_{3}^{*} & = & \frac{\mu_{3,0}}{\mu_{0,1}}-\frac{2\mu_{2,0}\mu_{1,1}+\mu_{1,0}\mu_{2,1}}{\mu_{0,1}{}^{2}}+\frac{2\mu_{1,0}\mu_{1,1}{}^{2}}{\mu_{0,1}{}^{3}}\label{eq:first-four-strongly intensive cumulants}\\
\kappa_{4}^{*} & = & \frac{\mu_{4,0}}{\mu_{0,1}}-\frac{3\mu_{3,0}\mu_{1,1}+\mu_{1,0}\mu_{3,1}}{\mu_{0,1}{}^{2}}-\frac{3\mu_{2,0}\mu_{2,1}}{\mu_{0,1}{}^{2}}\nonumber \\
 &  & +\frac{6\mu_{2,0}\mu_{1,1}{}^{2}+6\mu_{1,0}\mu_{2,1}\mu_{1,1}}{\mu_{0,1}{}^{3}}-\frac{6\mu_{1,0}\mu_{1,1}{}^{3}}{\mu_{0,1}{}^{4}}\nonumber 
\end{eqnarray}
which we expect will be most practical in common usage.

\section{Relation to Other Quantities\label{sec:Relation-to-Other}}

The name ``strongly intensive cumulants'' may seem contradictory,
as they are not cumulative like the standard cumulants. This name
was chosen due to their close relationship with cumulants. We'll work
with a fixed volume system so that there are no volume fluctuation
terms anywhere. Now consider the case where $X_{n}$ is completely
uncorrelated from $X_{1},X_{2},\dots,X_{n-1}$ at fixed volume such
that $\left\langle X_{1}^{r_{1}}X_{2}^{r_{2}}\cdots X_{n-1}^{r_{n-1}}\right\rangle \left\langle X_{n}^{r_{n}}\right\rangle =\left\langle X_{1}^{r_{1}}X_{2}^{r_{2}}\cdots X_{n}^{r_{n}}\right\rangle $
for any choice of $r_{1},r_{2},\dots,r_{n}$. It then follows that
\begin{eqnarray*}
D_{1}\left(\psi^{*}\right) & = & \frac{D_{1}\left(\phi\right)}{D_{n}\left(\phi\right)}=\frac{D_{1}\left(\phi\right)}{D_{n}\left(\left\langle e^{\xi_{i}X_{i}}\right\rangle _{X}\right)}=\frac{D_{1}\left(\phi\right)}{\left\langle X_{n}e^{\xi_{i}X_{i}}\right\rangle _{X}}\\
 & = & \frac{D_{1}\left(\phi\right)}{\left\langle X_{n}\right\rangle _{X}\left\langle e^{\xi_{i}X_{i}}\right\rangle _{X}}=\frac{D_{1}\left(\phi\right)}{\left\langle X_{n}\right\rangle _{X}\phi}\\
 & = & \frac{D_{1}\left(\ln\phi\right)}{\left\langle X_{n}\right\rangle _{X}}=\frac{D_{1}\left(\psi\right)}{\left\langle X_{n}\right\rangle _{X}}
\end{eqnarray*}
which directly implies
\begin{equation}
\kappa_{r_{1},r_{2},\dots r_{n}}^{*}=\frac{\kappa_{r_{1},r_{2},\dots r_{n}}}{\kappa_{0,0,\dots,0,1}}\label{eq:cumulant-correspondence}
\end{equation}
for $r_{1}>0$ and $n>1$. The definition of the integration constant
was, however, chosen such that this equality holds for $r_{1}=0$
as well. Thus, we find that the strongly intensive cumulants are equal
to their corresponding cumulants normalized by a volume term given
that $X_{n}$ is independent from the other components. For the case
of cumulant ratios, which are used frequently in experimental contexts,
this normalization cancels.

When we extend this to a variable volume, we find that the left hand
side of Eq. (\ref{eq:cumulant-correspondence}) does not change because
it is strongly intensive. This means that it corresponds to what the
cumulant ratio on the right hand side would be in the absence of volume
fluctuations. In a very real sense, this quantity corresponds to what
one would want to measure in the absence of volume fluctuations.

It is important to note here that this correspondence with the cumulants
depends only on the independence of $X_{n}$ at fixed volume and not
at all on the shape of its distribution. This means that any estimate
of the volume can be used for $X_{n}$, regardless of its noise profile.
Any independent quantity that has some dependence on the volume can
be used as $X_{n}$ if it is adjusted such that the dependence is
linearly proportional. This opens up the possibility of using volume
estimates for $\xi_{n}$ that have been previously considered too
noisy or irregular for volume determination. 

It is also worth mentioning that $\kappa_{r_{1},r_{2},\dots r_{n}}^{*}$
should equal zero when $X_{n}$ is chosen to be independent at fixed
volume, given that at least one of $r_{1},r_{2},\dots,r_{n-1}$ are
nonzero. This follows trivially from the cumulant correspondence demonstrated
in Eq. (\ref{eq:cumulant-correspondence}) because all joint cumulants
involving $X_{n}$ must be zero in order for it to be entirely independent.
This means that in experimental contexts the strongly intensive cumulants
with $r_{n}=0$ will likely be the most interesting. It also suggests
that evaluating how far the strongly intensive cumulants with $r_{n}>0$
are from zero might be useful for evaluating systematic errors involving
the independence of $X_{n}$.

The second-order strongly intensive cumulants are also closely related
to the $\Sigma$ and $\Delta$ fluctuation measures that were mentioned
earlier \citep{merick-1992,merik=2011,merick-2013}. Namely,
\begin{eqnarray*}
\Sigma\left[X_{1},X_{2}\right] & = & \kappa_{2}^{*}\left[X_{1},X_{2}\right]+\kappa_{2}^{*}\left[X_{2},X_{1}\right]\\
\Delta\left[X_{1},X_{2}\right] & = & \kappa_{2}^{*}\left[X_{1},X_{2}\right]-\kappa_{2}^{*}\left[X_{2},X_{1}\right]
\end{eqnarray*}
when the normalization factor on $\Sigma$ and $\Delta$ is chosen
to be $\nicefrac{\mu_{1,0}}{\mu_{0,1}^{2}}$ and the $\left[X_{1},X_{2}\right]$
denotes the order of the random vector components. The strongly intensive
cumulants have the same properties as these quantities but make the
relationship to the underlying physics more explicit.

\section{A Practical Example}

We consider here a practical example from heavy-ion physics. Higher
order fluctuations of conserved quantities in nuclear collisions has
been a topic of considerable interest in recent years. The most discussed
of these quantities has been the ratio of $\kappa_{4}/\kappa_{2}$
for net proton fluctuations as measured by the STAR experiment. With
respect to these measurements, it has frequently been reported that
``there are interesting trends, including e.g. the drop in the kurtosis
of the net-proton distribution at $\sqrt{s_{NN}}$= 27 and 19.6 GeV''
\citep{uli-trends}. We will study simulated nuclear collisions to
quantify the magnitude of bias introduced by volume fluctuations and
to demonstrate the utility of the strongly intensive cumulants for
addressing this issue.

Our analysis was run over 35 million Au+Au events at $\sqrt{s_{NN}}=7.7$
GeV that were generated using the UrQMD model \citep{urqmd-1998,urqmd-1999}.
The procedure followed was designed to closely mirror those used by
STAR \citep{net_charge_star,net_proton_star}. The net proton number
in each event was quantified as the number of protons minus the number
of antiprotons with transverse momenta in the range $0.4\ \text{GeV}<p_{T}<0.8\ \text{GeV}$
and pseudorapidities between $-0.5<\eta<0.5$. The detector efficiency
for these particles was assumed to be unity in order to disentangle
volume fluctuations from other effects. The approximate volume of
the system was measured using the multiplicity of charges particles
with $0.5<\left|\eta\right|<1.0$ and $p_{T}>0.15\ \text{GeV}$, a
quantity called $\text{Ref}_{\text{mult}:2}$ within STAR. This is
the centrality quantity used in the net charge, rather than net proton,
analysis at STAR but it was chosen to allow for a measurable $X_{n}$
variable.

Centrality was defined as the integrated percentile of $\text{Ref}_{\text{mult}:2}$,
from the largest values to the smallest. The cumulant ratios were
computed for each individual value of $\text{Ref}_{\text{mult}:2}$
and then averaged across each centrality bin. This was done to minimize
the impact of binning on the results. This procedure was repeated
three times with different binomial efficiencies for $\text{Ref}_{\text{mult}:2}$
each time: $p=\nicefrac{1}{3}$, $p=\nicefrac{2}{3}$, and $p=1$.
The efficiency of $p=\nicefrac{2}{3}$ is the most realistic but the
other two were included to demonstrate the effect of better and worse
volume resolutions. Additionally, the analysis was repeated with a
multiplicity variable that counted all pions and kaons produced in
each event. The results obtained using this multiplicity variable
represent an ideal case of resolution where the biases induced by
volume fluctuations should be largely eliminated.

\begin{figure}
\begin{centering}
\includegraphics[width=8cm]{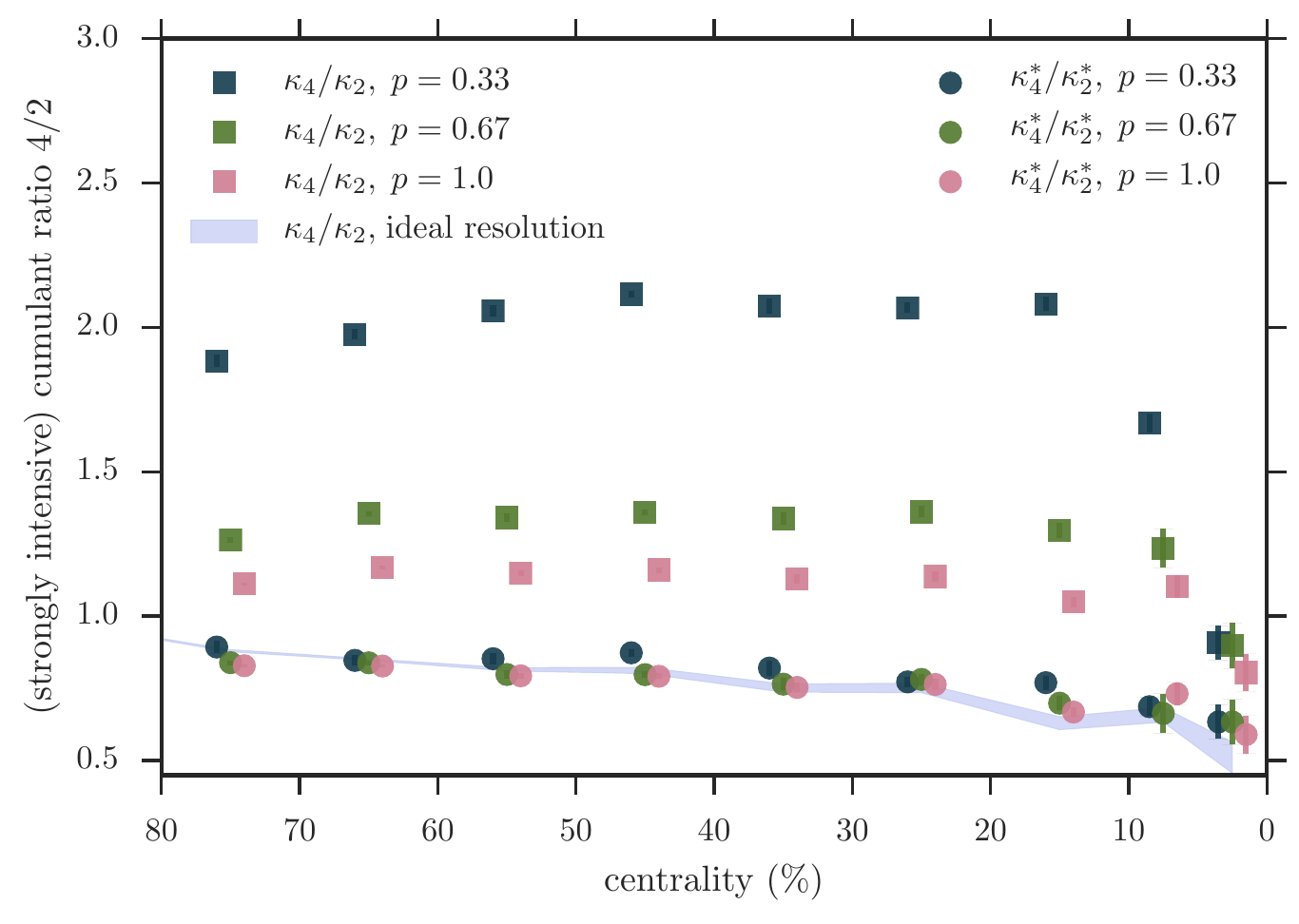}
\par\end{centering}

\protect\caption{Net Proton Cumulant Ratios\label{fig:Net-Proton-Cumulant}}

Standard and strongly intensive cumulant ratios of the net proton
distribution in UrQMD generated Au+Au events at $\sqrt{s_{NN}}=7.7$
GeV as a function of centrality. The square markers indicate standard
cumulant ratios and the circle markers indicate the strongly intensive
cumulant ratios for various centrality resolutions. The points are
offset slightly along the $x-$axis for clarity. The blue band indicates
what the standard cumulant ratios would be with an ideal centrality
resolution. We see that the strongly intensive cumulants correspond
very closely with the ideal centrality resolution scenario regardless
of the actual centrality resolution. The standard cumulant ratios,
on the other hand, are very significantly biased and depend quite
strongly on the centrality resolution. This shows very clearly that
the strongly intensive cumulants more accurately measure the desired
cumulant ratios in the inevitable presence of volume fluctuations.
\end{figure}

Finally, the analysis was repeated using strongly intensive cumulants
rather than standard cumulants. The $X_{1}$ component corresponded
again to the net proton number while $X_{2}$ was chosen to be the
number of charged pions and kaons with $0.4\ \text{GeV}<p_{T}<0.8\ \text{GeV}$
and $-0.5<\eta<0.5$ . The results can be seen in Fig. \ref{fig:Net-Proton-Cumulant}.

The most striking feature of the results is how dramatically the cumulant
measurements are shifted by volume fluctuations, in some cases by
well over a factor of two. The features observed in the STAR data
are on order of 20-40\% which can easily fall within the unquantified
systematic errors caused by volume fluctuations. This is particularly
true considering that $\sqrt{s_{NN}}$= 27 and 19.6 GeV were run with
different detector configurations than the other Beam Energy Scan
(BES) energies and therefore have significantly different systematics. 

Another interesting thing to note is the drastic shift in the cumulant
ratios in the most central events. This same trend can be seen in
the data and it can be explained by a suppression of volume fluctuations
when selecting on the highest multiplicity events. For mid-central
events, a single multiplicity value corresponds to high multiplicity
fluctuations from smaller volumes and low multiplicity fluctuations
from larger volumes. This results in a wide range of possible volumes.
For the highest multiplicity values there are only upward fluctuations
due to there being roughly a maximum volume attainable in a collision.
This leads to a tighter constraint of the volume near this maximum
value for the most central events. This, at first glance, might suggest
that the most central collisions are the most trustworthy. Unfortunately,
this is not the case in practice because these same bins are the most
sensitive to pileup events and secondary collision background. 

Moving on to the strongly intensive cumulants we find that they are
nearly identical for each of the resolution settings. Some small differences
can be seen due to physics correlations between the protons in $X_{1}$
and the pions and kaons in $X_{2}$, but, even still, the effects
of volume fluctuations are suppressed by well over an order of magnitude
compared to the standard cumulant ratios. The remaining differences
could be largely eliminated by a more careful choice of $X_{2}$.
These quantities aren't only invariant, they can also be seen to match
almost exactly with what the cumulant ratios would be in the case
of ideal centrality resolution. This very clearly illustrates both
the invariance under volume fluctuations of the strongly intensive
cumulants as well as their correspondence to the cumulants in the
absence of volume fluctuations.

\section{Conclusions}

A new set of statistical moments called the strongly intensive cumulants
have been presented. These quantities have been shown to be invariant
under both convolution and mixing with distributions sharing identical
strongly intensive cumulants. This property allows them to be experimentally
determined in ensembles of physical systems where the distribution
over volume is unknown. We have studied a practical example from heavy-ion
physics where the measurement of cumulant ratios has been shown to
be extremely biased due to the uncertainties in constraining the system
volume. In this same example, the strongly intensive cumulant ratios
have been shown to be almost entirely independent of how well constrained
the volume is. Furthermore, they have been shown to correspond to
the cumulant values that would be measured if the volume could be
almost perfectly constrained. For these reasons, it is clear that
the strongly intensive cumulants are better suited for determining
underlying physics in systems with imperfectly constrained volumes.

\section*{Acknowledgements}

The author would like to thank Gary Westfall for providing the UrQMD
data used in this paper and for valuable discussion. This research
was funded in part by NSF grant NSF-0941373 and DOE grant DE-FG02-03ER41259.

\bibliographystyle{apsrev4-1}
\bibliography{strongly_intensive}

\end{document}